# Plasmon enhanced optical tweezers with gold-coated black silicon

**D.G. Kotsifaki,[1] M. Kandyla,[1,] and P.G. Lagoudakis[2]**

[1]*Theoretical and Physical Chemistry Institute, National Hellenic Research Foundation, 48 Vasileos Constantinou Avenue, 11635 Athens, Greece*

[2]*Department of Physics and Astronomy, University of Southampton, Southampton, SO17 1BJ, UK*

## Abstract

Plasmonic optical tweezers are a ubiquitous tool for the precise manipulation of nanoparticles and biomolecules at low photon flux, while femtosecond-laser optical tweezers can probe the nonlinear optical properties of the trapped species with applications in biological diagnostics. In order to adopt plasmonic optical tweezers in real-world applications, it is essential to develop large-scale fabrication processes without compromising the trapping efficiency. Here, we develop a novel platform for continuous wave (CW) and femtosecond plasmonic optical tweezers, based on gold-coated black silicon. In contrast with traditional lithographic methods, the fabrication method relies on simple, single-step, maskless tabletop laser processing of silicon in water that facilitates scalability. Gold-coated black silicon supports repeatable trapping efficiencies comparable to the highest ones reported to date. From a more fundamental aspect, a plasmon-mediated efficiency enhancement is a resonant effect, and therefore, dependent on the wavelength of the trapping beam. Surprisingly, a wavelength characterization of plasmon-enhanced trapping efficiencies has evaded the literature. Here, we exploit the repeatability of the recorded trapping efficiency, offered by the gold-coated black silicon platform, and perform a wavelength-dependent characterization of the trapping process, revealing the resonant character of the trapping efficiency maxima. Gold-coated black silicon is a promising platform for large-scale parallel trapping applications that will broaden the range of optical manipulation in nanoengineering, biology, and the study of collective biophotonic effects.

[.] Corresponding author: kandyla@eie.gr

**Introduction**

Recent progress in the field of nanoscience has prompted the need for precise and non-invasive manipulation of individual nanoparticles. Among the possible approaches, optical trapping techniques, implemented by the use of optical tweezers, have emerged as a promising avenue to achieve this goal.[1-5] Conventional, far-field optical tweezers[3-5] have been successful in manipulating particles bigger than the wavelength of the trapping laser beam, but suffer from limitations when employed with nanometer-size particles, due to the diffraction limit of the focused trapping laser spot size.[1,2] The gradient radiation force, exerted on a trapped particle smaller than the wavelength of the trapping laser beam, is proportional to the cube of the particle radius,[4] hence the optical trapping efficiency decreases abruptly as the size of the trapped particle becomes smaller. Therefore, efficient and precise trapping of nanometer-size particles remains a challenge and requires high laser powers, when implemented with conventional optical tweezers. High incident photon flux results in photodamage of sensitive biological samples, such as DNA or proteins, which lose their viability during optical trapping.[6]

To overcome the aforementioned limitations, near-field optical trapping has been developed recently, based on plasmonic trapping substrates.[7-13] Plasmonic substrates allow for deep sub-wavelength confinement of light and yield enhanced electromagnetic near fields, resulting in efficient nanoparticle trapping with relatively low incident laser powers. Various substrates, consisting of metallic nanopillars,[9,14,15] nanoapertures,[11,12,16] and nanoantennas[10,17] have been employed, among others, for the implementation of plasmonic optical tweezers. Optical trapping forces on the order of nN have been obtained with plasmonic optical tweezers,[9,16] which are stronger than the trapping forces obtained with conventional optical tweezers for particles with the same polarizability.[18-20]

The use of plasmonic optical tweezers provides a set of control parameters, such as the distance of the trapped object from the substrate or the trapping laser wavelength, which affect the trapping process significantly and allow for customizing the optical trap for specific applications. A few experimental[21] and theoretical[22,23] studies have investigated the dependence of the radiation force, exerted on trapped dielectric particles, on the separation distance from a plasmonic substrate, which is expected to follow an exponential decay. Although the radiation force was found to decrease with increasing the separation distance from the substrate, the size of the trapped particles was bigger than the trapping beam wavelength, resulting in a non-exponential decay, mainly due to multiple scattering effects and propagation of the coupled evanescent field inside the particles. On the other hand, an important attribute of plasmon-mediated optical tweezers that is still evading a thorough spectroscopic investigation, is the wavelength-dependence of the trapping efficiency with respect to the plasmon-resonance. As with all resonant effects, such an investigation is essential for conclusively identifying the nature of the effect and distinguishing it from other light scattering or interference processes. Furthermore, the trapping force in plasmonic tweezers is expected to maximize when the wavelength of the trapping laser is resonant with the plasmon mode supported by the substrate, offering yet one more degree of freedom for optimizing the trapping efficiency.

In this work, we present a plasmonic optical tweezer implemented with femtosecond-laser nanostructured silicon substrates, also known as black silicon.[24-26] In contrast with traditional lithographic methods, the fabrication method employed here relies on simple, single-step, maskless tabletop laser processing of silicon in water, which is amenable to scalable fabrication. Coating the substrates with a thin layer of gold results in the spontaneous formation of gold nanoparticles on the surface. A thin intermediate layer of copper is employed for efficient thermal transport away from the trapping volume, in order to reduce bubble formation, which often compromises the functionality of plasmonic optical traps. The optical trapping setup operates either with a continuous wave (CW) infrared laser or with wavelength-tunable femtosecond laser systems (750 – 1300 nm). We observe significant enhancement of the trapping efficiency for 400-nm (diameter) polystyrene beads near the plasmonic substrate, to a maximum average value of 0.147, which is comparable to the highest trapping efficiencies reported to date, for trapped beads of similar sizes. Optical trapping force measurements versus the separation distance from the plasmonic substrate show an exponential decay of the trapping efficiency, indicating the trapping process is mediated by an evanescent electromagnetic field. The latter decay law is conclusively attributed to the plasmon-field of the substrate through a systematic wavelength-dependent characterization of the trapping efficiency that reveals the existence of a broad but clearly resolved resonance at $975 \pm 30$ nm. Even though an exponential decay of the optical trapping force away from similar substrates for a single trapping wavelength has been described in our previous work,[27] there were no conclusive evidence available regarding the plasmonic nature of the process, due to lack of spectroscopic measurements. In this work, the exponential decay with the distance from the substrate is observed repeatedly for all accessible trapping laser wavelengths, below and above the resonance wavelength. Furthermore, these results indicate that as the trapping wavelength approaches the plasmon resonance of the substrate, the exponential decay of the trapping efficiency with the distance from the substrate becomes steeper.

**Results**

**Substrate morphology**

Figure 1a shows a scanning electron microscope (SEM) image of a femtosecond-laser nanostructured silicon sample before metallic layer deposition (see Methods for details). The structuring process results in a quasi-ordered array of columnar nanopillars on the silicon surface, with a mean tip diameter of 160 nm and a mean height of 287 nm. The mean distance between neighboring pillars is 210 nm center-to-center. Irradiation of silicon by a train of femtosecond laser pulses in a liquid environment is known to generate nanopillars on the surface of the material due to ultrafast melting and interference effects,[24,28] resulting to what is known as black silicon. The inset in Fig. 1a shows a magnified SEM image of nanostructured silicon after Cu/Au deposition, where we observe that a layer of aggregated metallic nanoparticles is formed on the nanostructured silicon surface upon coating. The formation of metallic nanoparticles, instead of a smooth metallic film, is favored by the nanometric roughness of the $SiO_2$ surface layer of the laser-structured nanopillars.[29]

**Optical trapping: distance dependence**

We performed optical trapping measurements comparing the coated (3-nm Cu/50-nm Au) and uncoated, flat and nanostructured silicon substrates, employing the CW trapping laser. For a reference, we also used conventional glass substrates. The experimental setup is shown schematically in Fig. 1b (see Methods for details). Figure 2a shows the trapping efficiency, $Q$, as a function of the relative distance, $z$, between the trapping laser beam focus and the different substrates. The relative distance, $z$, is determined by the method described in Ref. 27. For a given distance $z$, each value of the trapping efficiency is calculated according to the definition $Q = Fc/nP$ (1), where $F$ is the optical trapping force, $P$ is the trapping laser power, $c$ is the speed of light, and $n$ is the refractive index of the surrounding medium. The optical forces were measured by the escape velocity method.[27] Each force value results from the average of five to ten independent measurements and each trapping efficiency value from three to nine laser power levels. Because the optical force depends on the trapping laser power, we express the experimental results in terms of trapping efficiency values, in order to decouple the findings from laser power variations. Plotting the trapping efficiency as a function of the distance from the substrate indicates not only how the magnitude of the trapping force changes as we approach each substrate but also how the performance of the optical trap, which depends on how strongly the trap immobilizes a particle for a given trapping laser power, changes for each substrate.

For all trapping measurements we avoided convective effects by keeping the trapping laser power low enough, to ensure a linear relation between $F$ and $P$ and to prevent bubble formation in the surrounding medium. Heating and thermal convection is pronounced in plasmonic trapping systems, due to Ohmic losses in the metallic substrates.[30-32] We followed the design of Wang *et al.*,[15] according to which combining a silicon substrate with a Cu/Au thin-film coating provides efficient thermal transport away from the trapping volume and minimizes convective effects. We confirmed that the Cu/Au-coated nanostructured silicon substrate prevents the formation of bubbles in the surrounding medium for the laser powers employed in this study, as opposed to an Ag-coated nanostructured silicon substrate (not shown here). Intense bubble formation in the deionized water containing the polystyrene nanobeads during optical trapping above the Ag-coated substrate prevented measurements near the surface[27] (see Supplementary movie).

In Fig. 2a we observe that for a distance $z = 1$ $\mu$m, the Cu/Au-coated nanostructured silicon substrate shows one order of magnitude enhancement of the trapping efficiency, compared with the other substrates. Furthermore, the trapping efficiency decays exponentially with increasing distance from the substrate, with a decay length of 575 nm. The solid line in Fig. 2a is a fit to the data according to the equation $Q = Q_o + Ae^{-bz}$ (2), where $Q_o$ is the trapping efficiency value at the distance $z = 10$ $\mu$m above the coated nanostructured silicon substrate and $A$, $b$ are fitting parameters. We do not observe an exponential decay of the trapping efficiency for the other substrates. The exponential decay of the trapping efficiency versus distance from the Cu/Au-coated nanostructured silicon substrate is attributed to the decay of the evanescent plasmon field of the substrate. In this context, the radiation force on the trapped beads increases near the substrate, due to the excitation of surface plasmon modes by the incident trapping laser beam, which explains the enhancement of the trapping efficiency compared to the other substrates. We fabricate a Cu/Au-coated flat silicon substrate and observe the trapping efficiency does not decay with increasing the distance from this substrate, excluding interpretations of the observed phenomena based on reflection, scattering, and/or interference effects of the trapping laser beam from the substrates.

**Optical trapping: wavelength dependence**

To further investigate the origin of the enhancement of the trapping efficiency for the coated nanostructured silicon substrate, we study the dependence of the trapping efficiency on the wavelength of the trapping laser beam. For this purpose, we employ a femtosecond laser trapping system that offers extended wavelength tunability and allows us to eliminate optical interference effects. The CW and femtosecond laser systems may be used interchangeably for optical trapping measurements for similar power levels and beam profiles.[33-35] Figure 2b shows the trapping efficiency, $Q$, as a function of the relative distance, $z$, between the trapping laser beam focus and the coated nanostructured silicon substrate for the three laser systems employed in this work, for the common wavelength of 1070 nm. The trapping efficiency values obtained with the femtosecond laser system, either alone (140 fs) or combined with the OPO (200 fs), start to deviate from each other and from the values obtained with the CW trapping laser at small distances from the substrate. We attribute this deviation to differences in the beam profile and mode quality of each laser system.[33]

Figure 3 shows the trapping efficiency as a function of the trapping laser wavelength, $\lambda$, at distances (a) $z = 1$ μm, (b) $z = 4$ μm, (c) $z = 6$ μm, (d) $z = 8$ μm, and (e) $z = 10$ μm above the coated nanostructured silicon substrate. For each distance, the data are fitted with a Gaussian function, represented by the solid curve in Figs. 3a – e. The femtosecond Ti:sapphire oscillator/OPO was employed for the wavelength range 750 – 1300 nm. For all distances, we observe the existence of a resonance wavelength at 975 ± 30 nm, where the trapping efficiency is highest, whilst the maximum value of the trapping efficiency decreases exponentially with the distance from the substrate. Indeed, Fig. 3f shows the maximum trapping efficiency, $Q_{max}$, resulting from the fits presented in Figs. 3a – e, as a function of the relative distance, $z$, above the coated nanostructured silicon substrate. The solid line represents an exponential fit to the data, according to the equation $Q_{\max} = Q_{\max,o} + Ae^{-bz}$ (3), where $Q_{max,o}$ is the average trapping efficiency value for the uncoated nanostructured silicon substrate for $\lambda = 1070$ nm and $A,\ b$ are fitting parameters. The overall maximum average trapping efficiency value we obtain in this study is $Q = 0.147 \pm 0.044$ for the wavelength of 1000 nm, at a distance $z = 1$ μm above the coated nanostructured silicon substrate. For the same conditions, we obtain the overall maximum optical trapping force and trapping efficiency values, which are $F = 21.96 \pm 0.64$ pN and $Q = 0.2028 \pm 0.0032$, respectively, for a trapping laser power of 24.5 mW. At $z = 1$ μm we could not take measurements for all experimentally accessible wavelengths, due to intense convection in the deionized water containing the trapped beads, even for low trapping laser powers. The average trapping efficiency value of 0.147 is comparable to the highest trapping efficiency values reported to date for similar trapped bead sizes, which include $Q = 0.27$ for 500-nm polystyrene beads trapped at a distance of 25 nm above gold bowtie nanoanntena arrays[36] and $Q = 0.1 \pm 0.02$ for 200-nm polystyrene beads trapped near gold nanopillar substrates.[9]

The exponential decay of the trapping efficiency with the distance from the coated nanostructured silicon substrate, which was observed with the CW laser at 1070 nm (Fig. 2a), is also repeatedly observed for all trapping laser wavelengths obtained with the femtosecond laser system. Figure 4 shows a surface plot of the trapping efficiency as a function of the relative distance between the trapping laser beam focus and the coated nanostructured silicon substrate and as a function of the trapping wavelength, below and above the resonance wavelength of $975 \pm 30$ nm. The experimental trapping efficiency values are interpolated on a color map surface to illustrate the decaying profile of the trapping efficiency versus the distance from the substrate and the resonant character of the observed enhancement. In Supplementary Fig. S2, the experimental data are plotted and fitted *vs*. the distance from the substrate, for each trapping wavelength separately.

**Discussion**

The wavelength resonance of the trapping efficiency indicates a plasmon-mediated trapping process, via coupling of the trapping laser beam with the surface plasmon modes of the substrate. Even though several studies[1,2] observe the enhancement of the  trapping efficiency for plasmonic substrates of various geometries, to the best of our knowledge the dependence of the trapping efficiency on the trapping laser wavelength has not been reported to date. The possibility that the wavelength dependence of the trapping efficiency is due to the wavelength dependence of the polarizability of the trapped particles, which affects the trapping force, is excluded from the analytic dependence of the trapping efficiency on the trapping laser wavelength in a conventional optical trap (see Supplementary Info). In this case, the trapping efficiency decreases monotonically with the wavelength of the trapping laser beam and does not present the resonant behavior, shown in Fig. 3. Also, the absorption depth of the trapping laser beam does not present a resonant behavior. The absorption depth of the trapping laser beam in gold varies between 12 – 18 nm for the present experimental conditions (aqueous surrounding medium, trapping wavelengths 750 – 1300 nm), which is lower than the nominal thickness of the gold layer we deposit (50 nm), for all wavelengths. Since the absorption depth is also a monotonically decreasing function of the trapping beam wavelength, it cannot explain the resonant behavior we observe in Fig. 3.

The plasmonic optical tweezers presented in this work are based on a single-step, tabletop laser-processing fabrication method in water, which does not require the use of vacuum or lithographic masks, making the process simple and cost-effective. Furthermore, there is no inherent limitation on the size of the laser-processed silicon area. This paves the way for large-scale fabrication of plasmonic tweezer systems. Moreover, in this work the trapping process does not require special alignment of the trapped object on the plasmonic substrate x-y plane, since the enhancement of the trapping force and trapping efficiency do not rely on specific hot spots on the substrate. The laser fabrication method provides a relatively uniform nanostructured pattern on the silicon surface. Indeed, similar substrates have been found to exhibit spatially uniform SERS enhancement factors.[29] The spontaneous formation of metallic nanoparticles on the nanostructured substrate during thermal evaporation simplifies further the fabrication process. The morphology of laser-fabricated silicon structures can be controlled primarily via the laser pulse duration in combination with the surrounding medium. Irradiation of silicon by femtosecond, picosecond, or nanosecond laser pulses in a gas environment results in the formation of micropillars,[13,37,38] while irradiation by femtosecond laser pulses in a liquid results in the formation of nanopillars.[24,28,39] For a given pulse duration and surrounding medium, the morphology can be further controlled via the incident laser fluence, laser wavelength, and number of pulses.[40] The size and density of the metal nanoparticles depends on the amount (nominal thickness) of the metal coating, the deposition rate, and the substrate temperature.[29] Post-deposition laser heating and melting of the metallic layer can lead to additional modifications of the nanoparticle morphology.[41] Tuning the fabrication parameters allows for controlling the morphology and optical properties of the plasmonic substrates, including the resonance wavelength, and maintaining the repeatability in the measurement of the trapping efficiencies observed here.

In summary, we have demonstrated a plasmonic optical tweezer system based on simple femtosecond-laser structuring of silicon in water and subsequent coating with a Cu/Au layer. The substrate morphology creates a strong plasmonic optical trap. The trapping efficiency decays exponentially with the distance from the substrate, following the exponential tail of the local electromagnetic field. A resonance wavelength appears at $975 \pm 30$ nm, for which the trapping efficiency maximizes for all distances from the substrate. The maximum average trapping efficiency we obtain is $Q = 0.147 \pm 0.044$ for 400-nm polystyrene beads, which is comparable to the highest trapping efficiencies reported to date for similar trapped bead sizes. The plasmonic substrate shows efficient thermal transport away from the trapping volume and allows for measurements near the surface, enhancing the functionality of the optical trap. The proposed scheme of femtosecond laser optical trapping, integrated with plasmonic substrates, paves the way for novel applications, including enhancing the diagnostic potential of optical tweezers. Femtosecond plasmonic optical tweezers may benefit the study of live fluorescent cells or the detection of viruses due to low operating laser power levels[10] and the capability of probing the nonlinear optical response of the specimen. These studies, which are beyond the scope of the current paper, will be the subject of a future investigation. The ease, cost-effectiveness, and scalability of the fabrication process offers an attractive system for large-scale parallel trapping applications, which will broaden the range of optical manipulation in nanoengineering and nanobiology and will enable novel many-body physics as well as collective biophotonic effects to be investigated, among others.

**Methods**

**Fabrication process**

In order to fabricate the plasmonic substrates, we employed an amplified femtosecond laser for nanostructuring the surface of silicon wafers. A regenerative Ti:sapphire amplifier was used to generate 800-nm center wavelength, 180-fs pulses at a repetition rate of 20 KHz. The pulses were frequency-doubled to a center wavelength of 400 nm using a $BBO_3$ crystal. The average power of the pulse train after the $BBO_3$ crystal was 25 mW and the pulse train was focused on a silicon wafer surface, producing a 20-$\mu$m FWHM beam diameter. The silicon wafer was placed in a cuvette filled with distilled water and irradiated by approximately 60k pulses. The femtosecond-laser structured silicon wafers were coated by thermal evaporation with a nominal 3-nm copper layer followed by a 50-nm gold layer (Cu/Au), resulting in a dispersion of gold nanoparticles on the surface, providing efficient thermal management.[27] Uncoated femtosecond-laser nanostructured silicon wafers were used as reference trapping substrates.

**Description of trapping setup**

For optical trapping measurements, we employed a home-built trapping setup,[42] operating with three different trapping laser systems: 1) a CW near-infrared fiber laser with a wavelength of 1070 nm, 2) a femtosecond Ti:sapphire oscillator, producing 140-fs pulses at a repetition rate of 80 MHz, and 3) an optical parametric oscillator (OPO), producing 200-fs pulses. The setup is shown schematically in Fig. 1b. The trapping laser beam is focused by a high numerical aperture, oil-immersion microscope objective lens (NA = 1.4, x63) near the plasmonic substrates. Fluorescent polystyrene beads of 400-nm diameter (Duke Scientific) are trapped in deionized water near the focal point of the trapping laser beam. The bead solutions are contained in a homemade cell, consisting of a cover glass slide, the plasmonic substrate and spacers.  In order to excite a fluorescence signal from the trapped beads and image them on a CCD camera, we use a CW probe laser beam ($\lambda = 473$ nm, $P_{max} = 1.32$ mW after the objective lens) that is also shone through the objective lens on the bead solution. For escape velocity measurements, the plasmonic substrates are placed on a piezo-controlled stage with a position resolution of 50 nm. During measurements, we systematically verified that the trapped beads were not mechanically pinned on the substrates and they were released when switching off the trapping laser. Also, we did not observe photobleaching effects during optical trapping measurements. Measurements were performed on random locations of the plasmonic substrates.

**Acknowledgements**

Financial support of this work by the General Secretariat for Research and Technology, Greece, (project Polynano-Kripis 447963) is gratefully acknowledged.

**Author contributions**

D.G.K. developed the concept and carried out the optical trapping experiments, performed the analysis of data and wrote the manuscript. M.K. fabricated the plasmonic substrates. M.K. and P.G.L. conceived the experiments. All authors contributed to the analysis of the results and commented on the manuscript.

**Competing financial interests:** The authors declare no competing financial interests.

**<u>Figure Captions</u>**

**Figure 1: Experimental design.** (a) Schematic diagram of the optical trapping setup. (b) Scanning electron microscope (SEM) image, at side (45°) view of the surface, of an uncoated femtosecond-laser nanostructured silicon substrate. Inset: magnified view of a femtosecond-laser nanostructured silicon substrate after coating with 3 nm of copper and 50 nm of gold.

**Figure 2: Trapping efficiency measurements versus separation distance.** (a) Trapping efficiency, $Q$, as a function of the relative distance, $z$, between the trapping laser beam focus and various substrates. Relative distance is the measured distance with respect to the zero position. Solid line: exponential fit to the data obtained with the nanostructured silicon substrate, coated with Cu/Au. Trapping data were obtained with the CW laser setup. (b) Trapping efficiency, $Q$, as a function of the relative distance, $z$, between the trapping laser beam focus and the nanostructured silicon substrate, coated with Cu/Au, obtained with three different laser systems for the same wavelength (1070 nm). The laser systems are: CW laser at 1070 nm; femtosecond laser at 1070 nm (140 fs); optical parametric oscillator (OPO) at 1070 nm (200 fs). The x-error corresponds to the Rayleigh length of the trapping laser beam and the y-error to the standard deviation of the trapping efficiency measurement.

**Figure 3: Wavelength - dependent characterization of trapping efficiency**. Trapping efficiency, $Q$, as a function of the trapping laser wavelength, measured at distances (a) $z = 1$ μm, (b) $z = 4$ μm, (c) $z = 6$ μm, (d) $z = 8$ μm, and (e) $z = 10$ μm above the nanostructured silicon substrate, coated with Cu/Au. Solid lines: Gaussian fit to the data. Trapping data were obtained with the femtosecond laser setup. The y-error corresponds to the standard deviation of the trapping efficiency measurement. (f) Maximum trapping efficiency, $Q_{max}$, resulting from the fits presented in Figs. 3a – e, as a function of the relative distance, $z$, between the trapping laser beam focus and the substrate. Relative distance is the measured distance with respect to the zero position. Solid line: exponential fit to the data. The y-error corresponds to the error of the Gaussian fit.

**Figure 4: Surface plot of the trapping efficiency**, $Q$, as a function of the relative distance, $z$, between the trapping laser beam focus and the Cu/Au-coated nanostructured silicon substrate and as a function of the trapping wavelength, obtained with the femtosecond laser setup. Relative distance is the measured distance with respect to the zero position. The experimental trapping efficiency values are interpolated on a color map surface. The colored bar indicates the minimum and maximum interpolated trapping efficiency values.

# Figures

## Fig. 1

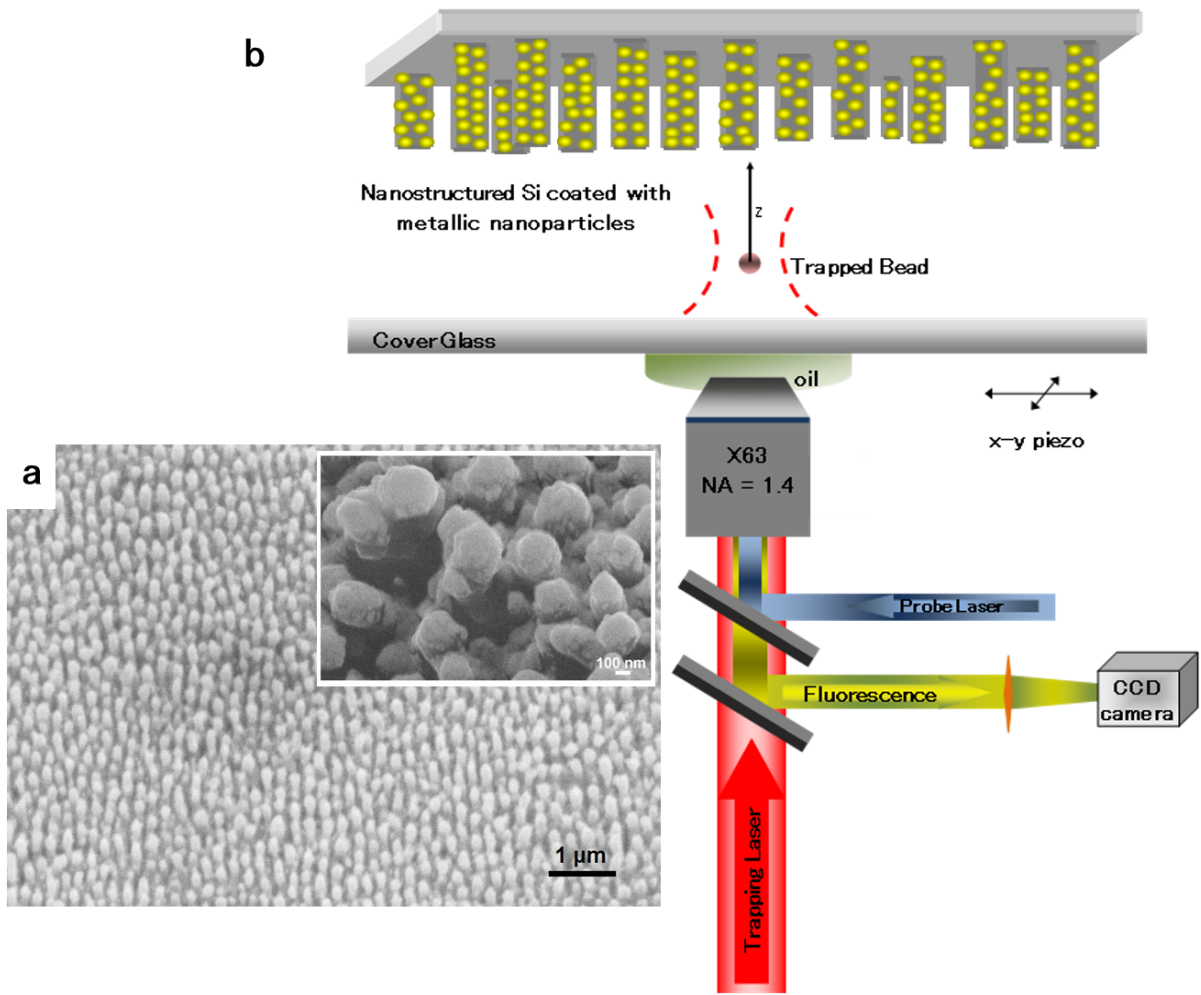

**Fig. 2**

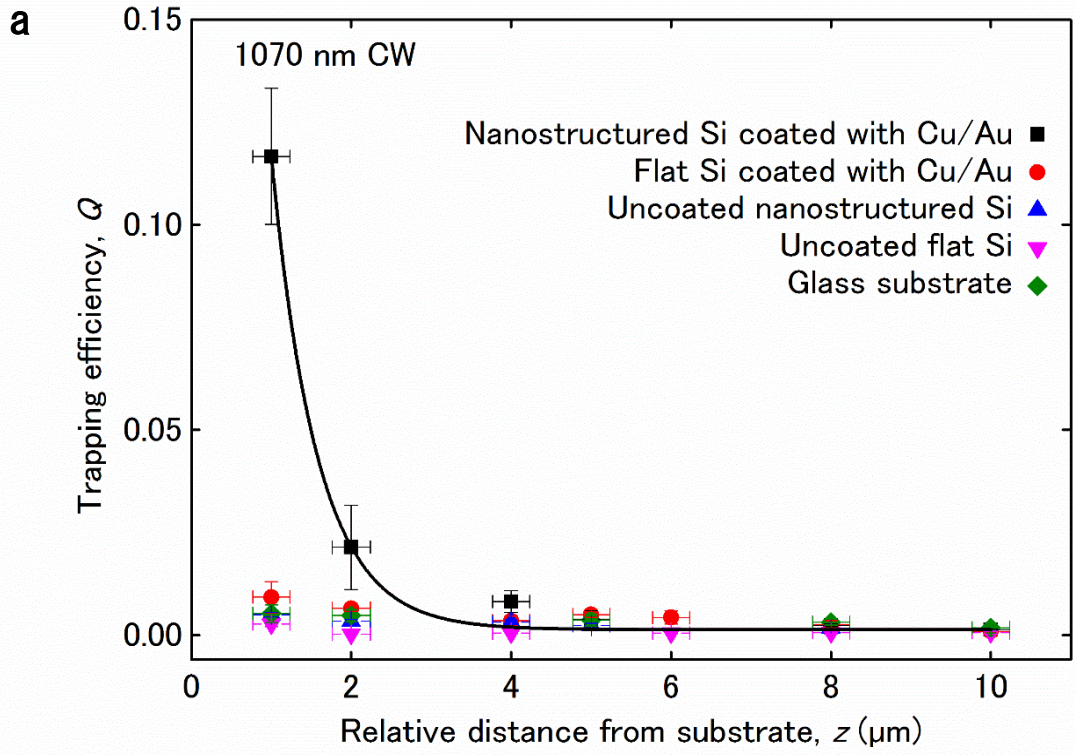

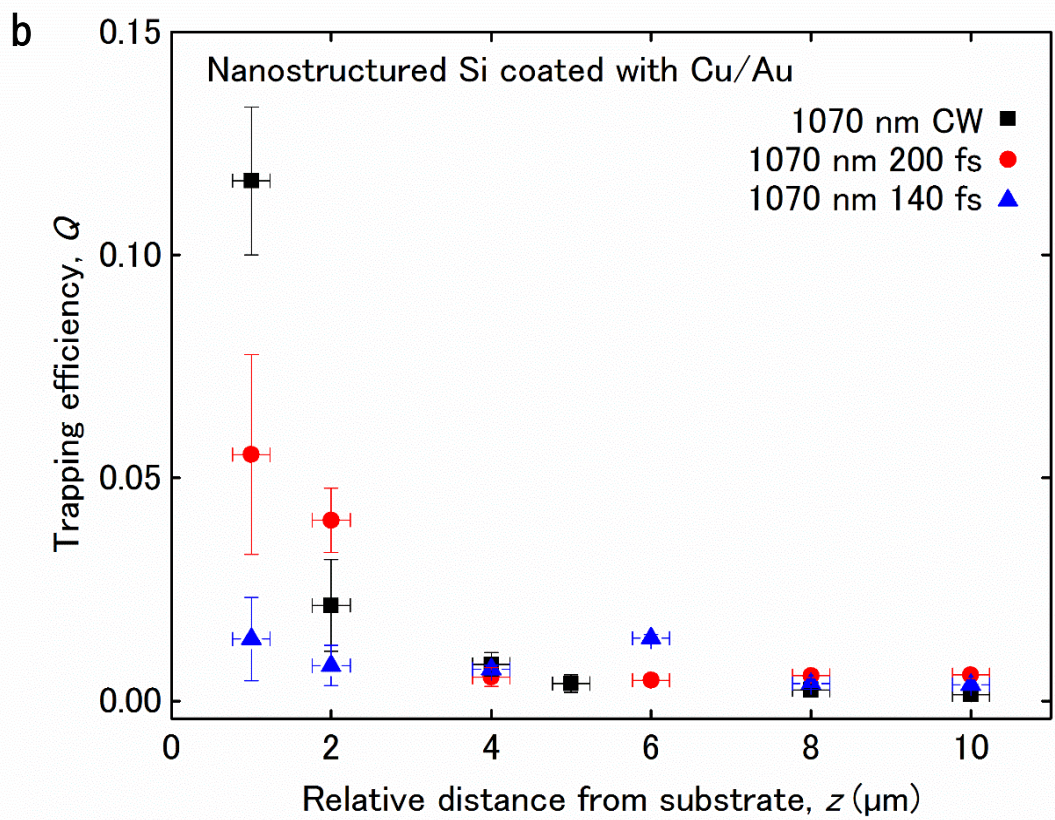

**Fig. 3**

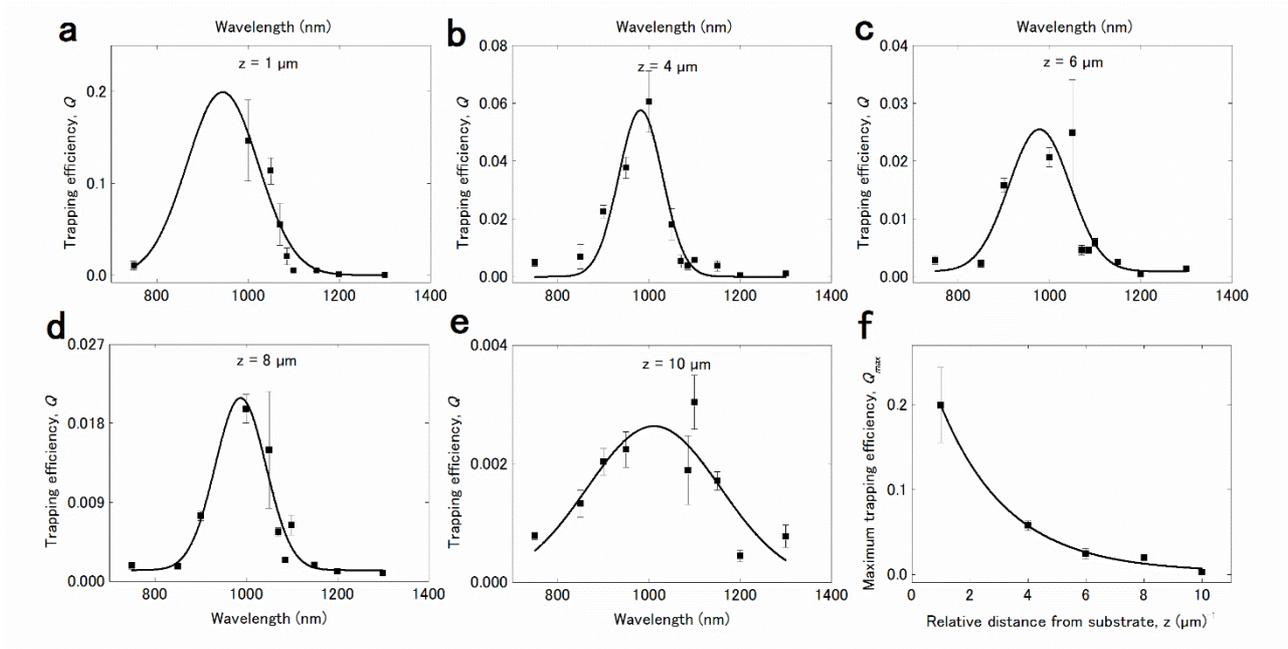

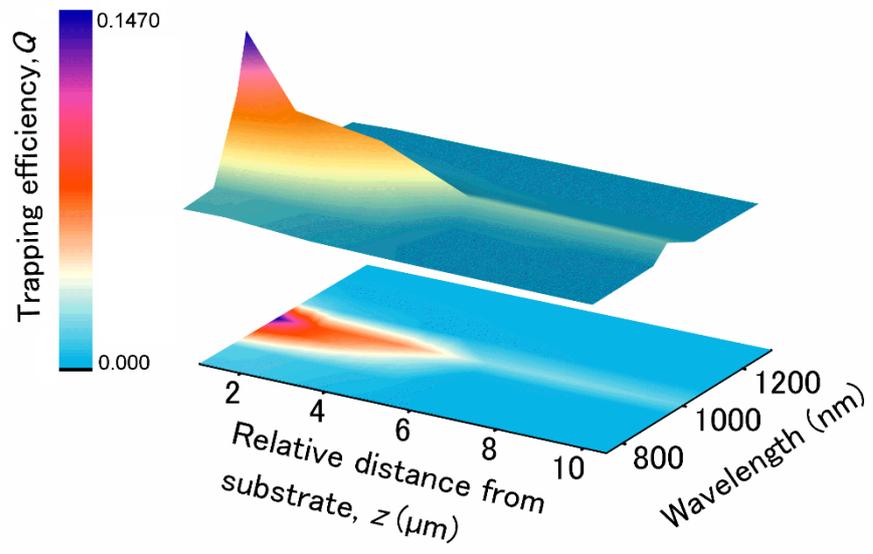

**Plasmon enhanced optical tweezers with gold-coated black silicon**

*Domna G. Kotsifaki, Maria Kandyla,[1] and Pavlos G. Lagoudakis*

**Supplementary movie**

The movie shows bubble formation on an Ag-coated nanostructured silicon substrate.

**Effect of polarizability**

In order to investigate the effect of the trapped particle polarizability on the wavelength dependence of the trapping efficiency in a conventional optical trap, we derive an analytical expression of the latter using the Rayleigh approximation[S1]:

$$Q = 4\pi^3 N A^3 e^{-1/2} \left( \frac{r}{\lambda} \right)^3 \left( \frac{\varepsilon_p - \varepsilon_m}{\varepsilon_p + 2\varepsilon_m} \right) \quad \text{(Eq. S1)}$$

where $\lambda$ is the wavelength in the surrounding medium ($\lambda = \lambda_o/n_m$, $\lambda_o$ the wavelength in vacuum and $n_m$ the refractive index of the surrounding medium), NA the numerical aperture of the objective lens, $r$ the particle radius, and $\varepsilon_p$, $\varepsilon_m$ the dielectric permittivity of the particle and the surrounding medium, respectively. Eq. S1 has been derived by employing the definition $Q = Fc/n_mP$, where $c$ is the speed of light and $P$ the power of the trapping laser beam. The trapping force, $F$, is calculated according to Eq. 17 in Ref. S1, for $(x, y, z) = (w_0/2, 0, 0)$ for which the gradient force takes its maximum value. The beam radius at the beam-waist position, $w_0$, is calculated as $w_0 = \lambda/(\pi \text{ NA})$.

Figure S1 shows the theoretical trapping efficiency for 400-nm diameter polystyrene beads in deionized water, as a function of the incident trapping laser wavelength. NA = 1.4 has been used in the calculations, which corresponds to the experimental numerical aperture. The trapping efficiency decreases monotonically with the wavelength of the trapping laser beam and does not present a resonant behavior, as the one shown in Fig. 3 of the main paper.

---

[1] Corresponding author: kandyla@eie.gr

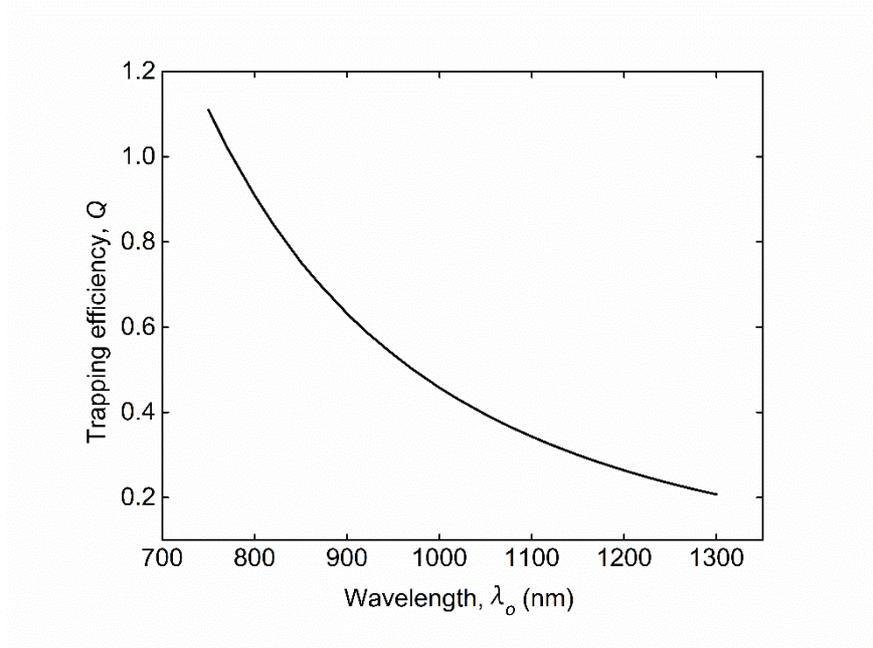

**Figure S1:** Theoretical trapping efficiency, $Q$, as a function of the trapping laser wavelength, $\lambda_o$.

**Exponential decay of trapping efficiency**

Figure S2 shows a semilog plot of the trapping efficiency as a function of the relative distance between the trapping laser beam focus and the coated nanostructured silicon substrate, for several trapping wavelengths, below and above the resonance wavelength of $975 \pm 30$ nm. The data for wavelengths 750 nm, 850 nm, 950 nm, and 1000 nm are offset for clarity purposes. We fit the trapping efficiency data for each trapping laser wavelength with an exponential function, according to the equation $Q = Q_o + Ae^{-bz}$, where $Q_o$ is the experimental trapping efficiency value at a distance $z = 10$ μm above the coated nanostructured silicon substrate for each trapping laser wavelength and $A$, $b$ are fitting parameters. The fitting results are shown in Fig. S2 as solid lines, which indicate that as the trapping wavelength approaches the plasmon resonance of the substrate, the exponential decay of the trapping efficiency with the distance from the substrate becomes steeper.

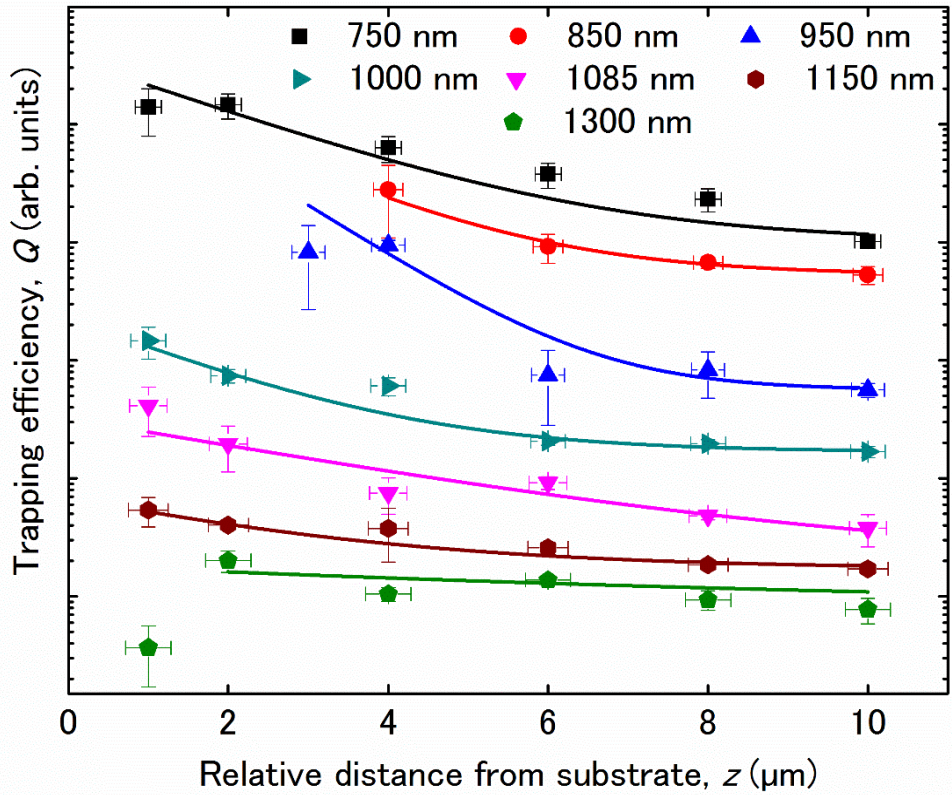

**Figure S2:** Semilog plot of the trapping efficiency, $Q$, as a function of the relative distance, $z$, between the trapping laser beam focus and the Cu/Au-coated nanostructured silicon substrate, for various trapping laser wavelengths, obtained with the femtosecond laser setup. Relative distance is the measured distance with respect to the zero position. Solid lines: exponential fits to the data. The data for wavelengths 750 nm, 850 nm, 950 nm, and 1000 nm are offset.